\documentclass[aps,prl,twocolumn,showpacs,preprintnumbers,amsmath,amssymb]{revtex4-1}

\usepackage{graphicx}

\begin{document}

\preprint{}

\title{Implementation of low-loss superinductances for quantum circuits}

\author{Nicholas A. Masluk}
  \email{nicholas.masluk@yale.edu}
\author{Ioan M. Pop}
\author{Archana Kamal}
\author{Zlatko K. Minev}
\author{Michel H. Devoret}
\affiliation{Department of Applied Physics, Yale University, 15 Prospect Street, New Haven, CT 06511}

\date{\today}

\begin{abstract}
The simultaneous suppression of charge fluctuations and offsets is crucial for preserving quantum coherence in devices exploiting large quantum fluctuations of the superconducting phase.  This requires an environment with both extremely low DC and high RF impedance.  Such an environment is provided by a superinductance, defined as a zero DC resistance inductance whose impedance exceeds the resistance quantum $R_Q = h/(2e)^2 \simeq 6.5\ \mathrm{k\Omega}$ at frequencies of interest (1 - 10 GHz).  In addition, the superinductance must have as little dissipation as possible, and possess a self-resonant frequency well above frequencies of interest.  The kinetic inductance of an array of Josephson junctions is an ideal candidate to implement the superinductance provided its phase slip rate is sufficiently low.  We successfully implemented such an array using large Josephson junctions ($E_J >> E_C$), and measured internal losses less than 20 ppm, self-resonant frequencies greater than 10 GHz, and phase slip rates less than 1 mHz.
\end{abstract}

\pacs{85.25.Cp, 74.81.Fa, 74.50.+r, 64.70.Tg}

\maketitle

The emerging field of quantum electronics exploiting large fluctuations of the superconducting phase is limited by the practical challenge of engineering an electromagnetic environment which suppresses simultaneously the quantum fluctuations of charge and the random low-frequency fluctuations of offset charges. The small value of the fine structure constant $\alpha=1/137$ entails a fundamental asymmetry between flux and charge quantum fluctuations, strongly favoring the latter. To illustrate this, let us consider the simplest case of a dissipationless LC oscillator, where the charge $Q$ on the capacitor plates and the generalized flux $\Phi$ across the inductor are conjugate variables. The ratio between quantum fluctuations of charge, $\delta q = \delta Q/(2e)$, and flux, $\delta \varphi = \delta \Phi / \Phi_0$, in the ground state of the oscillator is given by $\delta \varphi / \delta q = Z_{0}/R_{Q}$. Here $Z_{0}=\sqrt{L/C}$ is the characteristic impedance of the oscillator and $R_{Q}=h/(2e)^{2}=6.5\, k\Omega$ is the superconducting resistance quantum. Using only geometrical inductors and capacitors, the characteristic impedance of the oscillator $Z_{0}$ can not exceed the vacuum impedance $Z_{vac}=\sqrt{\mu_{0}/\epsilon_{0}}$, thus imposing quantum fluctuations of charge at least an order of magnitude larger than flux fluctuations: $\delta\varphi/\delta q<Z_{vac}/R_{Q}=8\alpha$.

In order to exceed the vacuum impedance and suppress quantum charge fluctuations, circuit elements with extremely high impedance are required. On chip resistors and long chains of Josephson junctions (JJs) in the dissipative regime have already been used to provide high impedance environments \cite{KUZMIN1991,Lotkhov2003,Corlevi2006}. However, these Ohmic components do not help shunting charge offsets and, being dissipative, they tend to destroy the quantum coherence of the devices. In order to simultaneously suppress quantum charge fluctuations and offsets, we need a circuit element which possesses three key attributes: high impedance at frequencies of interest, perfect conduction at DC and extremely low dissipation. These attributes define the so-called ``superinductance'' \cite{superinductance_Kitaev}.

\begin{figure}[htbp]
  \begin{center}
    \includegraphics[scale=0.3]{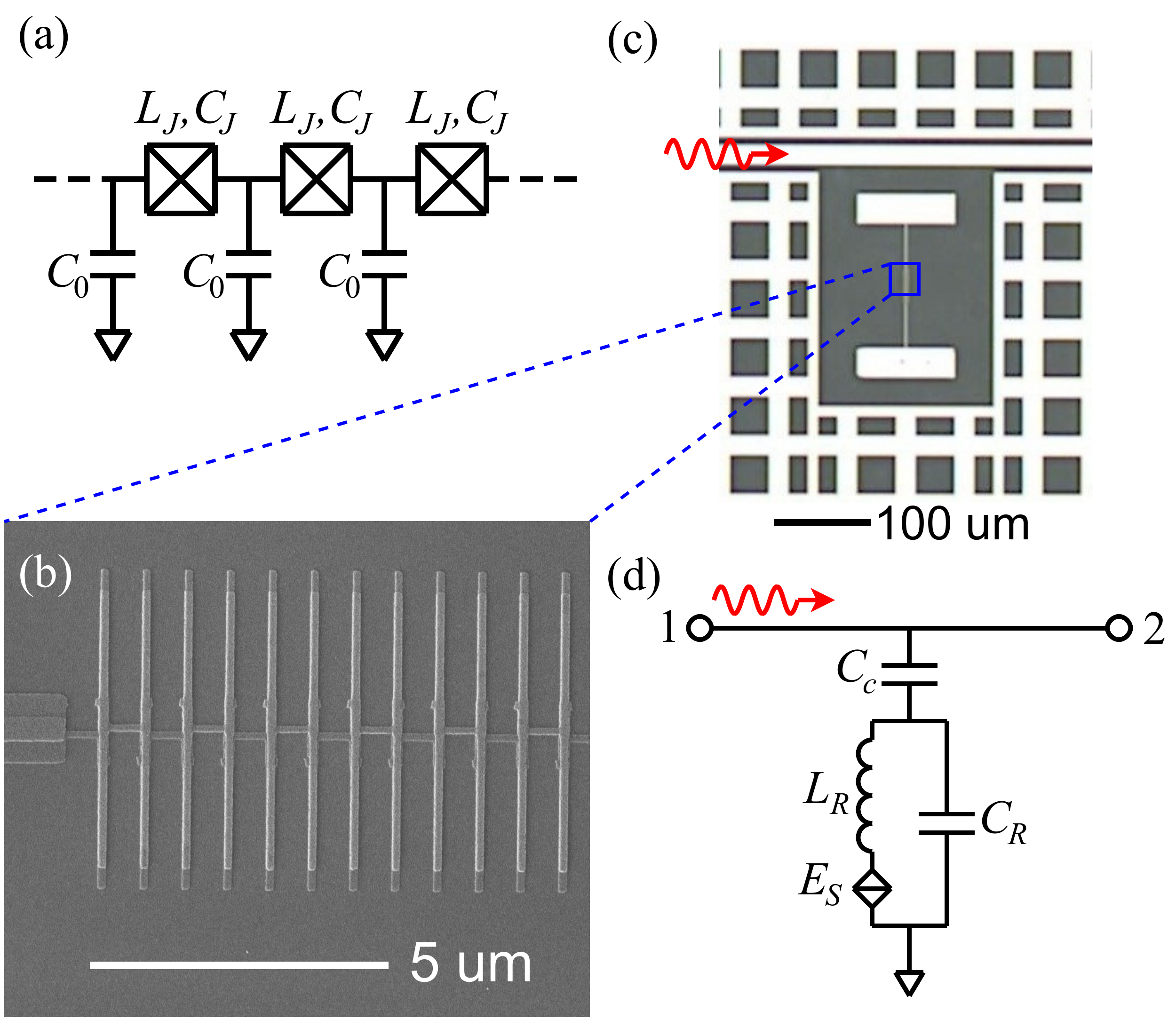}
    \caption{(a) Schematic representation of an array of Josephson junctions.  Capacitances of islands to ground and across tunnel junctions are denoted with $C_0$ and $C_J$, respectively.  The junction inductance is given by $L_J$.  (b) SEM image of the Josephson junction array fabricated using the bridge-free technique \cite{BFT, rigetti}.  (c) Optical image of an LC resonator.  The large pads implement the resonator capacitance as well as coupling capacitances to the CPW feedline.  An array of Josephson junctions between the pads implements the superinductance.  The ground plane is patterned with flux-trapping holes.  (d) Low-frequency model for the device shown in (c).  The phase slip element (split diamond) in series with the superinductance represents the collective contribution of phase slips through all junctions in the array.  The characteristic energy of the phase-slip element is denoted by $E_S$.}
    \label{fig:1} %Label must come before end of center, otherwise labeling will refer to an incorrect number, but after \caption{}
  \end{center}
\end{figure}

In this Letter, we report the successful implementation and detailed characterization of superinductances using the large kinetic inductance of arrays of Josephson junctions (see Fig~\ref{fig:1}(a)), following the proof-of-concept shown in the fluxonium circuit \cite{Manucharyan2009}. This first superinductance implementation suffered from coherent quantum phases-slips (CQPS) \cite{Manucharyan2012,Pop2012}, which constitute additional degrees of freedom, difficult to control experimentally. We show that we can completely suppress the CQPS by using large JJs with Josephson energy $E_{J}$$\backsimeq100\, E_{C}$, where $E_{C}=e^{2}/(2C_J)$ is the charging energy of one junction. 

Superconducting nanowires are also likely candidates for implementing superinductances \cite{Bezryadin2000,Mooij2006}. Unfortunately, they show significant internal dissipation which is not yet well understood, and are more challenging to fabricate. The JJ arrays may also suffer from dissipation, either due to coupling to internal degrees of freedom \cite{Fazio2001,Rastelli2012} or to a dissipative external bath \cite{Lobos2011}. Indeed, transport measurements on large arrays of JJ show the appearance of a superconducting to insulating transition (SIT) with decreasing Josephson energy $E_{J}$ \cite{Chow1998,Kuo2001,Takahide2006}. Previous implementations of resonators where the inductive energy is given by JJ arrays have yielded internal quality factors in the range of a few thousands \cite{Castellanos-Beltrana2007,Palacios-Laloy2008}. We show here that the problem of dissipation can be alleviated by using large JJs to realize superinductances far from the SIT region with internal quality factors an order of magnitude above the previously reported values.

The superinductances are formed by an array of closely spaced Josephson junctions on a C-plane sapphire substrate, as shown in Fig.~\ref{fig:1}.  We fabricated the junctions by e-beam lithography and double angle evaporation of aluminum using the bridge-free technique of \cite{BFT, rigetti}.  Prior to aluminum deposition, the substrate is cleaned of resist residues using an oxygen plasma \cite{pop}.  We minimize the width of the connecting wires between junctions in order to reduce parasitic capacitances to ground, which ultimately lower the self-resonant frequency of the superinductance \cite{hutter2011}.

We characterize our superinductances at low temperatures and microwave frequencies by incorporating them in lumped element LC resonators capacitively coupled to a co-planar waveguide in the hanger geometry.  The resonator response is measured in transmission.  An optical image of a typical device and its low-frequency circuit model are shown in Figs.~\ref{fig:1}(c,d).

The samples were mounted on the mixing chamber stage (15 mK) of a dilution refrigerator inside a copper box, enclosed in an aluminum-Cryoperm-aluminum shield with a Cryoperm cap.  We used  two 4-12 GHz Pamtech isolators and a 12 GHz K\&L multi-section low-pass filter before the HEMT amplifier, and installed copper powder filters on the input and output lines.

\begin{figure}[htbp]
  \begin{center}
    \includegraphics[scale=0.6]{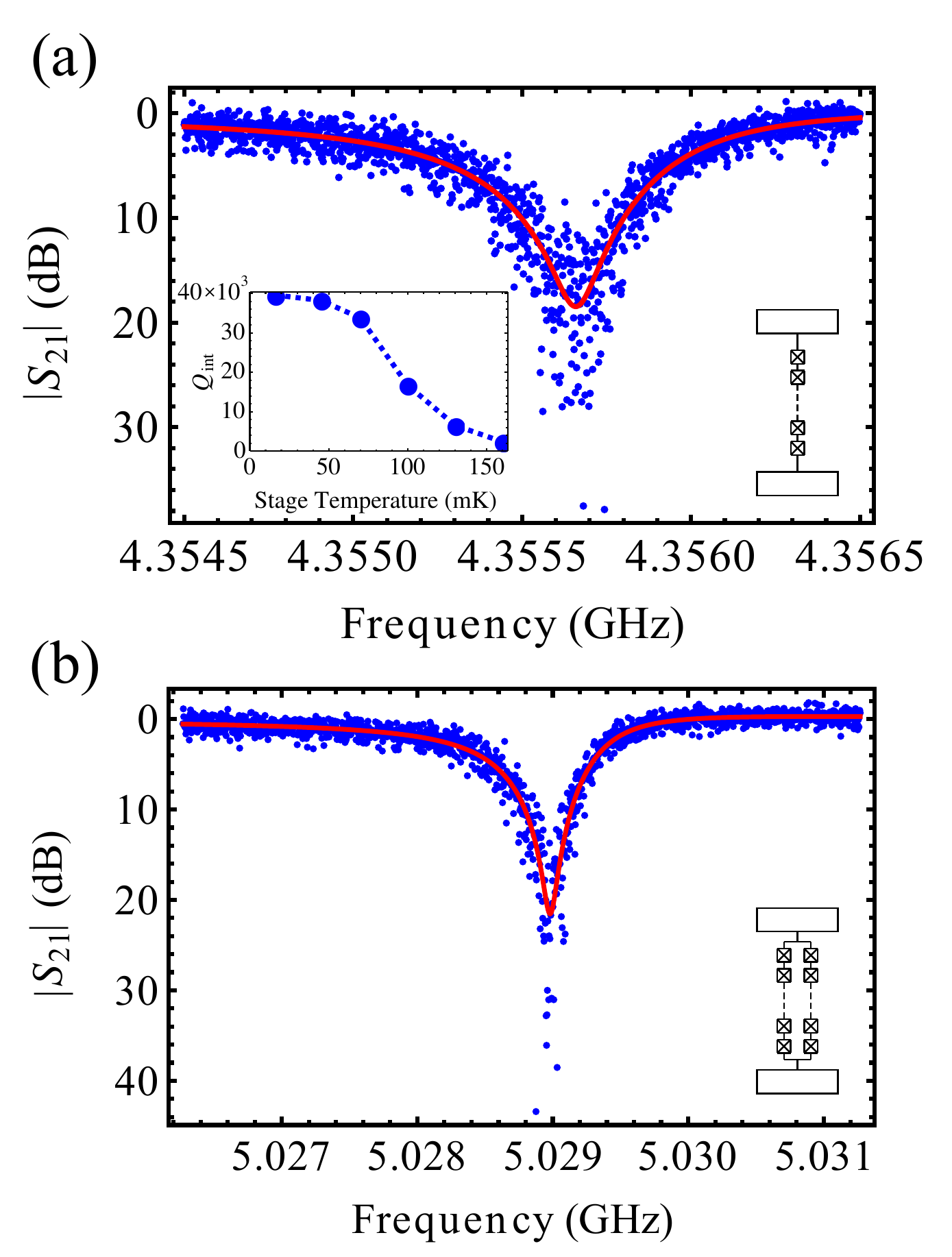}
    \caption{(a) Typical microwave transmission data for an 80-junction resonator measured with order one photon circulating power; solid line is theoretical prediction corresponding to $Q_{int} > 37,000$.  Inset:  Temperature dependence of the internal quality factor of the resonator.  Dashed line is a guide for the eye.  (b) Same measurement as in (a) for a 160-junction array loop; solid line corresponds to $Q_{int} > 56,000$.  Parameters for the low-frequency model in Fig.~\ref{fig:1}(d) are $C_c = 1.6$ fF, $C_R = 7.2$ fF and $L_R = 150$ nH for (a) and $C_c = 1.8$ fF, $C_R = 11$ fF and $L_R = 76$ nH for (b).}
    \label{fig:2}
  \end{center}
\end{figure}

The internal quality factor of a resonator is extracted by fitting the transmission data about the resonance with the response function \cite{geerlings}
\begin{equation}
  S_{21}(f) = 1 - \frac{Q_{ext}^{-1} - 2i \frac{\delta f}{f_R}}{Q_{tot}^{-1} + 2i \frac{f - f_R}{f_R}},
\end{equation}
where $Q_{tot}$ and $Q_{ext}$ are the total and external quality factors, $f_R$ is the resonant frequency, and $\delta f$ characterizes asymmetry in the transmission response profile.  The internal quality factor is given by $Q_{int} = \frac{Q_{ext}Q_{tot}}{Q_{ext}-Q_{tot}}$.  We show a typical measured response for a resonator with an 80-junction superinductance in Fig.~\ref{fig:2}(a).  This yields an internal quality factor of 37,000 for the resonator at 15 mK (stage temperature), corresponding to a loss in the superinductance of better than 27 ppm.  We note that an unknown portion of the internal loss comes from the capacitors.  The inset of Fig.~\ref{fig:2}(a) shows the dependence of the internal quality factor on the stage temperature.  A similar resonator with two 80-junction arrays in parallel was measured to have a quality factor of 56,000 (Fig.~\ref{fig:2}(b)), corresponding to a superinductance loss of better than 18 ppm.  The external quality factors, $Q_{ext}$, of both resonators were 5,000.

We next evaluate the self-resonant modes of an $N$-junction array. The Lagrangian of the array is
\begin{eqnarray}
    \mathcal{L} & = & \sum_{n=1}^{N} \frac{1}{2} C_{0} \dot{\Phi}_{n}^{2} + \frac{1}{2} C_{J} (\dot{\Phi}_{n} - \dot{\Phi}_{n+1})^{2} \nonumber\\
    & & \qquad - \frac{1}{2} \frac{(\Phi_{n} - \Phi_{n+1})^{2}}{L_{J0}},
\end{eqnarray}
where $\Phi_{n}$ are the node fluxes associated with each superconducting island.  We express each node flux as a superposition of discrete Fourier mode amplitudes,
\begin{eqnarray}
    \Phi_{n} = \frac{1}{\sqrt{N}}\sum_{k=1}^{N} e^{i \frac{\pi k}{N} n} \Phi_{k},
    \label{DFT}
\end{eqnarray}
which leads to a diagonal Hamiltonian of the following form:
\begin{eqnarray}
    \mathcal{H}
    &=& \sum_{k=-N/2}^{N/2} \mathbb{C}_{k k^{'}}^{-1} Q_{k}Q_{-k'} + \mathbb{L}_{k k^{'}}^{-1} \Phi_{k} \Phi_{-k^{'}}.
\end{eqnarray}

Here $Q_{k}$ are the canonical conjugate ``charge" variables to the fluxes $\Phi_{k}$, and
\begin{eqnarray}
    \mathbb{C}_{k k^{'}} &=& \delta_{k k^{'}} \left[\frac{C_{0}}{2} + C_{J} \left(1 - \cos \frac{\pi k}{N}\right)\right]; 
    \quad k \in\left[-\frac{N}{2}, \frac{N}{2}\right]\nonumber\\
    \\
    \mathbb{L}_{k k^{'}} &=& \delta_{k k^{'}} \frac{L_{J0}}{\left(1- \cos \frac{\pi k}{N}\right)},
    \label{LK}
\end{eqnarray}
are respective capacitance and inductance matrices in the Fourier basis. This immediately leads to a dispersion relation of the form,
\begin{eqnarray}
    \omega_{k} = (L_{kk}C_{kk})^{-1/2} 
    = \omega_{0} \sqrt{\frac{1-\cos \frac{\pi k}{N}}{\frac{C_{0}}{2 C_{J}} + \left(1 - \cos \frac{\pi k}{N}\right)}},
\label{linfreq}
\end{eqnarray}
where $\omega_0 = 1/\sqrt{L_J C_J}$ is the plasma frequency of a single junction.  The coupling capacitance pads at each end of the array (see Fig.~\ref{fig:1}(c)) load down the eigenfrequencies (see supplementary information for calculation details).

\begin{figure}[htbp]
  \begin{center}
    \includegraphics[scale=0.4]{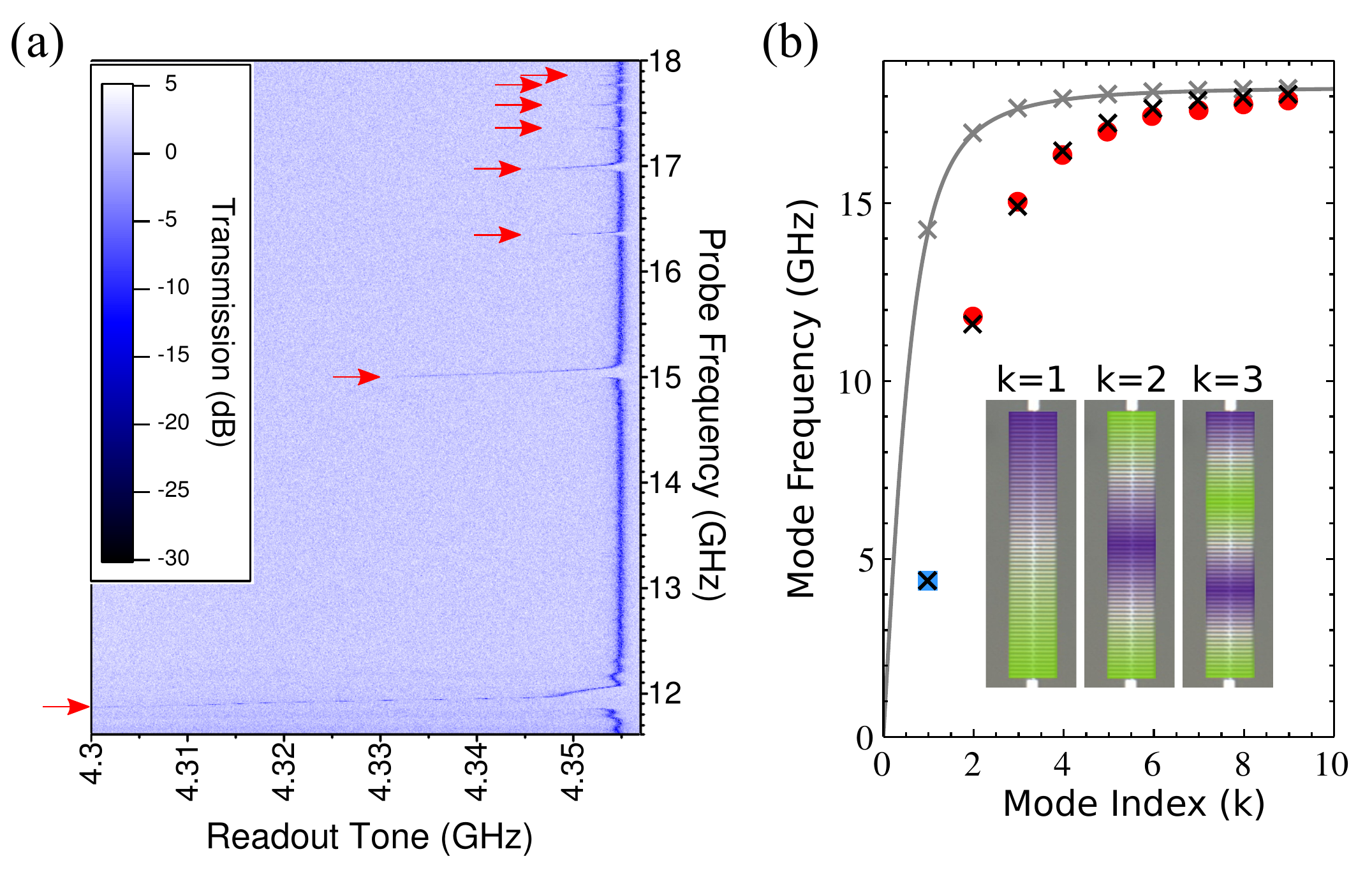}
    \caption{(a) Shift of the lowest frequency mode of the 80-junction resonator upon application of a probe tone.  Due to the weak non-linearity of the array (theory predicts of order 10 MHz/photon \cite{nonlinearity}), shifts occur when the probe tone matches a plasmonic resonance of the array (marked with red arrows).  (b) Measured plasma mode frequencies of the 80-junction array:  The blue square represents the fundamental resonant frequency of the resonator (4.355 GHz), and red circles indicate the plasma modes pointed out by red arrows in (a).  The black crosses denote the calculated plasma frequencies.  The gray curve represents the dispersion relation without including corrections due to the coupling capacitors (open boundary conditions), and markers show corresponding plasma frequencies.  The inset in (b) shows the voltage profile along the array for the first three plasma modes.}
    \label{fig:3}
  \end{center}
\end{figure}

The data presented in Fig.~\ref{fig:2} corresponds to measurements of the $k = 1$ mode.  The frequencies of higher modes, $k \ge 2$, of the array lie outside the band of our measurement setup.  In order to observe these modes we exploit their cross-Kerr interaction, which is induced by the junction nonlinearity.  This interaction leads to a frequency shift of the lowest mode ($k = 1$) when higher modes are excited.  In Fig.~\ref{fig:3}(a) we show the results of a two-tone measurement of an 80-junction resonator (same device as presented in Fig.~\ref{fig:2}(a)), where we continuously monitor the $k = 1$ mode while sweeping the frequency of a probe tone.  When the probe tone is resonant with a higher mode of the array, we observe a drop in the $k = 1$ frequency.

Fig.~\ref{fig:3}(b) shows the comparison between the measured frequencies of the array modes and the theoretically predicted values.  The black crosses show the renormalized mode frequencies calculated after incorporation of the corrections due to coupling capacitance pads (see supplementary information), which are in good agreement with the measured frequencies represented by colored markers. The parameters extracted from the fit were $C_{0} = 0.04$ fF, $C_J = 40$ fF and $L_{J} = 1.9$ nH, with a confidence range of 20\%. The simulated value of the capacitance to ground $C_{0}= 0.09$ fF agrees within a factor of 2 with the inferred value from the fit. Room temperature resistance measurements of the junction arrays yield a value of $L_{J} = 2.1$ nH, which agrees with the fit within 10\%.  Using the fit parameters we calculate the dispersion relation for the bare superinductance, shown as a gray curve in Fig.~\ref{fig:3}(b).  We note that the lowest resonant frequency of the bare superinductance is 14.2 GHz, which meets the design criterion of having self resonances well above the frequency range of interest (1 - 10 GHz).

\begin{figure}[htbp]
  \begin{center}
    \includegraphics[scale=0.3]{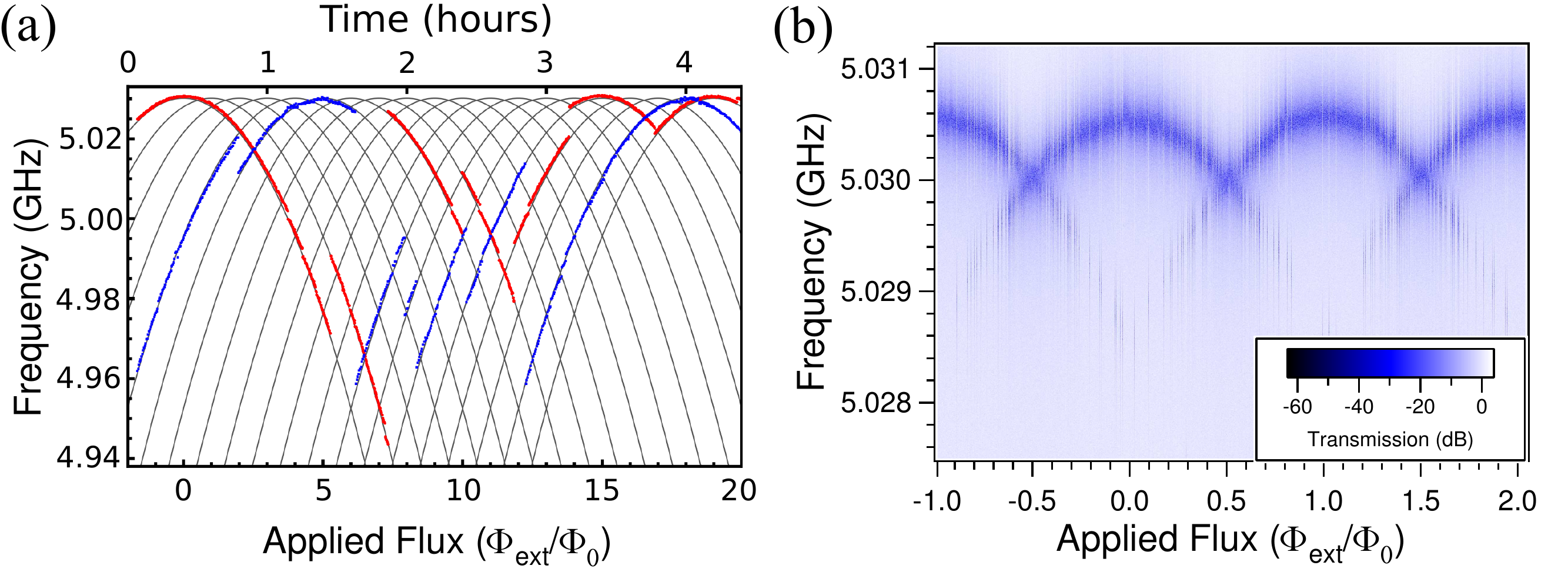}
    \caption{(a) Lowest mode frequency versus applied flux bias for resonator with 160-junction array loop.  Flux bias is swept up (red) and down (blue) over the course of several hours.  Gray curves represent quasi-classical predictions for resonator frequencies at different integer values of flux quanta in the array loop.  The only adjustable parameter is the number of junctions, found to be $150 \pm 10$.  (b) Same measurement as in (a) over a smaller flux range, but before each measurement a high power pulse at one of the plasma modes is applied in order to reset the resonator to the lowest flux state.}
    \label{fig:4}
  \end{center}
\end{figure}

In order to characterize the phase slip rate of the superinductances, we monitor the frequency of the $k = 1$ mode of a resonator formed by two superinductances in parallel, while sweeping an external magnetic field.  As flux bias is increased, the persistent current induced in the loop increases, and results in a drop in frequency of the mode (Fig.~\ref{fig:4}(a)).  This behavior can be described to lowest order in junction nonlinearity by the quasi-classical expression
\begin{equation}
  f(\Phi_{ext}) = \frac{f_{R}}{\sqrt{1 + \frac{1}{2} (\frac{2\pi}{N}(\frac{\Phi_{ext}}{\Phi_0} - m))^2 }},
\end{equation}
where $\Phi_{ext}$ is the applied flux bias, and $m$ is the integer number of flux quanta inside the loop.  A phase slip event is associated with an integer change in $m$, which we observe as a jump in resonator frequency.  As we increase the flux bias, the phase slip rate is enhanced and frequency jumps become more probable \cite{mlg}.  We swept the flux bias applied to the loop over several flux quanta before a phase slip event occurred.  The typical duration between phase slips recorded in this experiment was over an hour, Fig.~\ref{fig:4}(a).  This is a remarkably low phase slip rate of well under 1 mHz for a loop of 160 junctions, significantly lower than previously reported values on shorter arrays \cite{Manucharyan2012,Pop2010}.

Due to the extremely low phase slip rate, in order to measure the resonator in the lowest flux state, we employ an active resetting scheme.  We apply a high power pulse at one of the $k \ge 2$ modes before measuring the location of the lowest resonant frequency.  This high power pulse activates phase slips, allowing the loop to settle into the lowest flux state.  Using this protocol, we tracked the resonant frequency in the lowest flux state, as shown in Fig.~\ref{fig:4}(b).  Discrete inverted parabolas are observed, which allow us to unambiguously calibrate the number of flux quanta in the loop.  The fitted value for the total number of junctions differs by 6\% from the actual number, which can be explained by the classical nature of the theory which does not take quantum fluctuations into account.

In conclusion, we have demonstrated the successful implementation of superinductances using arrays of Josephson junctions, and performed the first detailed characterization of a superinductance.  We measured superinductances in the range of 100--300 nH, self resonant frequencies above 10 GHz, internal losses less than 20 ppm, and phase slip rates below 1 mHz.  The long lifetimes of persistent current states in the array loop also demonstrate the low DC resistance of the arrays.  With these parameters, the Josephson junction array superinductance enriches significantly the quantum electronics toolbox.  Applications would include further suppression of offset charges in superconducting qubits \cite{Manucharyan2009}, high Q and tunable lumped-element resonators, on-chip bias tees, kinetic inductance particle detectors \cite{Day2003}, and precise measurements of Bloch oscillations \cite{bloch}.

We would like to acknowledge fruitful discussions with Luigi Frunzio, Kurtis Geerlings, Leonid Glazman, Wiebke Guichard, Zaki Leghtas, Mazyar Mirrahimi, Michael Rooks and Rob Schoelkopf. Facilities use was supported by YINQE and NSF MRSEC DMR 1119826. This research was supported by IARPA under Grant No. W911NF-09-1-0369, ARO under Grant No. W911NF-09-1-0514 and NSF under Grant No. DMR-1006060.  As this manuscript was completed, we learned of a similar implementation of a superinductance using a different array topology \cite{Bell2012}.

\vfill
%-------------------------------------------------------------------------------
\pagebreak
\onecolumngrid

\setcounter{equation}{0}
\setcounter{figure}{0}

\renewcommand{\thefigure}{S\arabic{figure}}
\renewcommand{\theequation}{S\arabic{equation}}

\section{
Supplemental material for \\
  Implementation of superinductances for quantum circuits
}

\section{Simulation of Parasitic Capacitance in Junction Arrays}
The parasitic capacitance to ground $C_0$ of the islands in a Josephson junction (JJ) array lowers the self-resonant frequency of the superinductance. We optimize the array geometry in Ansoft Maxwell electromagnetic field simulation software in order to minimize the parasitic capacitance.

\begin{figure}[htbp]
  \begin{tabular}{c}
    \includegraphics[height=5cm]{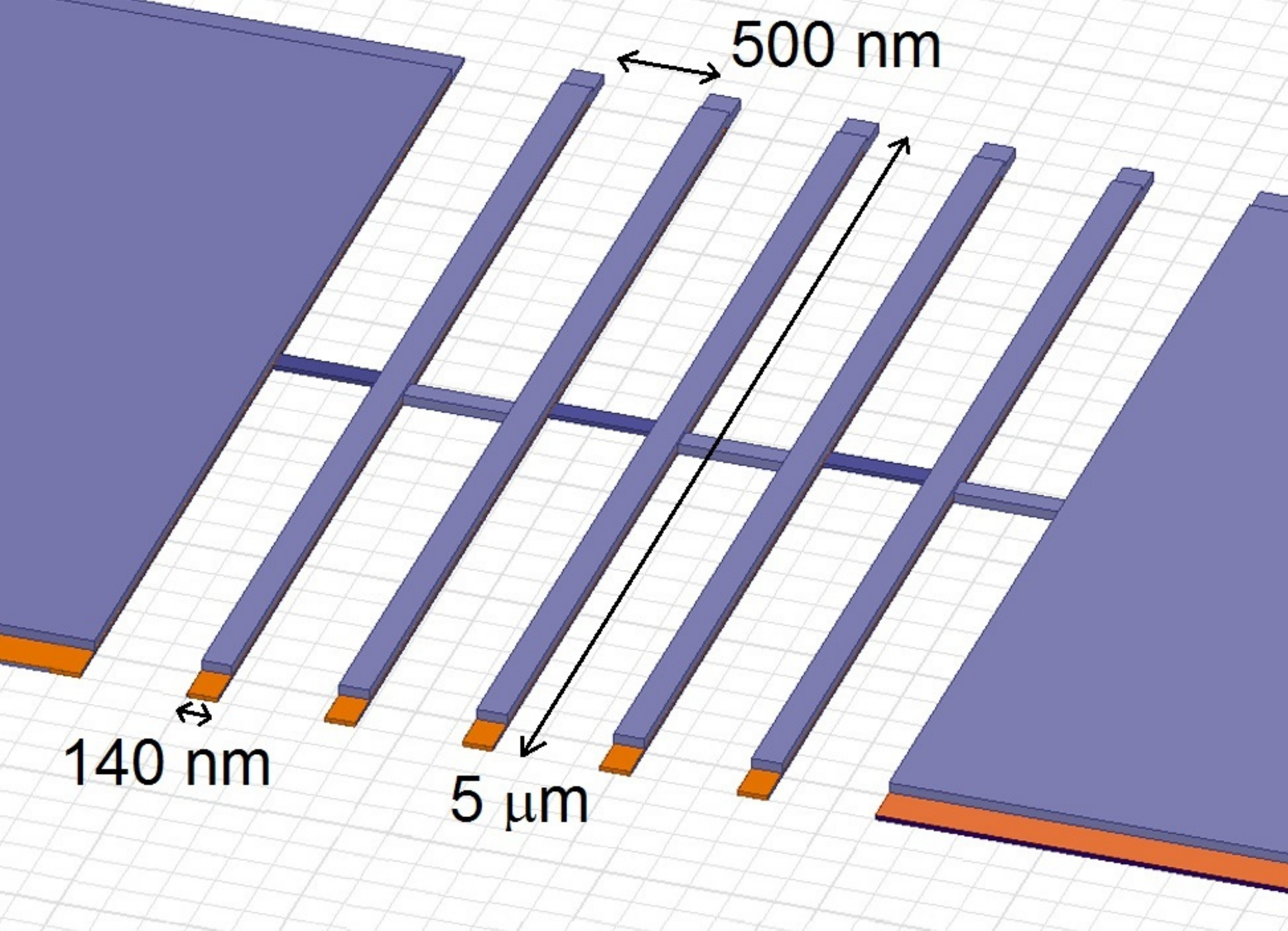} \\
    (a)
  \end{tabular}
  \begin{tabular}{c}
    \includegraphics[height=5cm]{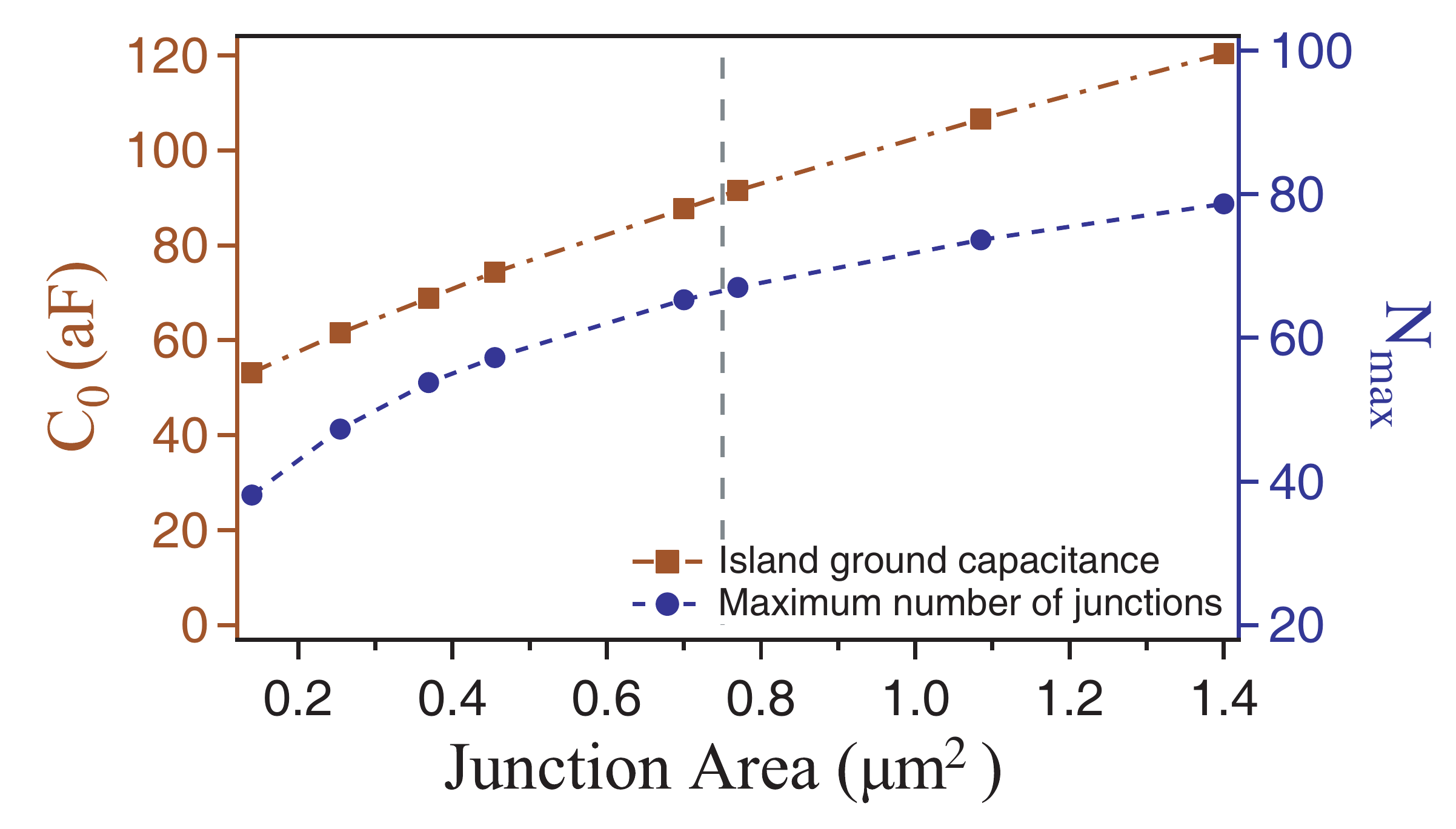} \\
    (b)
  \end{tabular}
  \caption{
    (a)
     Simplified BFT junction array model. The orange color represents the thin layer of dielectric (1 nm thick AlO$_x$ oxide), and purple represents the perfect conductor (superconducting aluminum).
    (b) Simulated parasitic capacitance to ground and inferred $N_{max}$ (Eq.~(\ref{equ:maxN})) for the design in (a).  The junction area is varied by sweeping its length.  The dashed vertical gray line indicates the parameters for the sample presented in the main text.  The lines joining points are a guide for the eye.
  }
  \label{fig:s1}
\end{figure}

Our model for the array includes the connection pads and substrate (Fig.~\ref{fig:s1}(a)). We model the junction oxide layer by a 1 nm thick dielectric material with relative permittivity of 6. Despite only simulating 5 junctions in the array, we can robustly obtain the parasitic capacitance for the full structure. Increasing the number of junctions in the model leads to no significant change in the computed value of the parasitic capacitance.

Numerical simulations revealed that junctions with large, 1:30, aspect ratios minimize the parasitic capacitance (Fig.~\ref{fig:s1}(b)). Fabricating junctions with such large aspect ratios using the Dolan bridge technique is limited by the collapse of the bridge.
% Fluxonium was 200nm by 2500nm or  12.5:1,  BFT is .14um by 5um is 35:1. can go up to 70:1 
The bridge-free technique (BFT) \cite{rigetti, BFT} allows the fabrication of arbitrary size and aspect ratio junctions. The BFT also minimizes the width of the connecting wires between junctions.  Our simulations show that this leads to a reduction of 60\% in parasitic capacitance as compared with the standard Dolan bridge technique.

The parasitic capacitance $C_0$ of each island limits the maximum number of junctions in the array to 
\begin{equation}
  N_{max}\simeq\pi\sqrt{\frac{C_J}{C_0}},
  \label{equ:maxN}
\end{equation}
a value such that the first self-resonant mode of the array lies $\sim$30\% below the plasma frequency of a single junction \cite{popThesis}.  In Fig.~\ref{fig:s1}(b) we present a plot of the simulated $C_0$ and $N_{max}$ as a function of JJ size for large aspect ratio junctions.  For our device we obtain $N_{max} = 70$.

\section{Calculating the Scattering Matrix for the Hanger Geometry}
\begin{figure}[htbp]
  \begin{center}
    \includegraphics[scale=0.5]{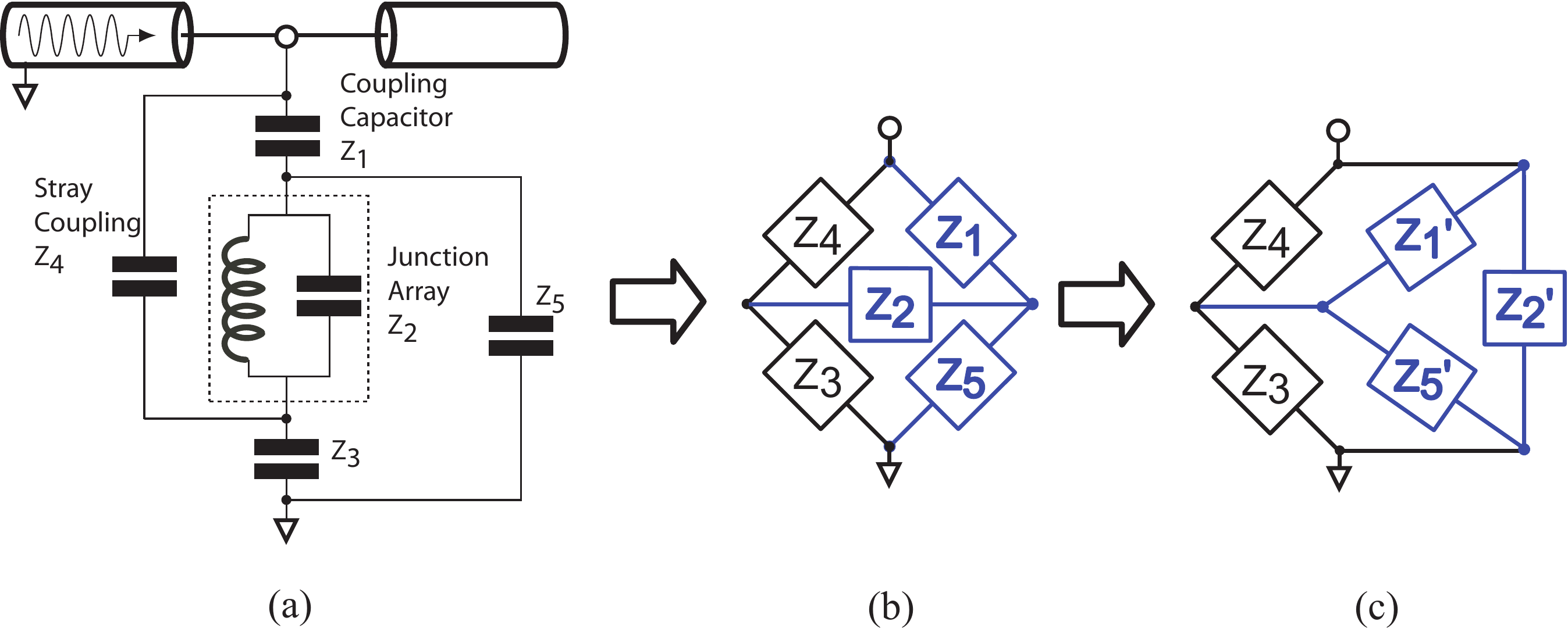}
    \caption{(a) Model of resonator coupled to a CPW in the hanger geometry. The schematic in (b) elucidates the Wheatstone bridge topology of (a).  The resultant circuit after a Y-$\Delta$ transform of the circuit in (b) is shown in (c).}
    \label{fig:s2} 
  \end{center}
\end{figure}

In this section we describe the calculation of scattering matrix of an array resonator coupled to a CPW in a hanger geometry.  The circuit model is shown in Fig.~\ref{fig:s2}(a), which corresponds to the device shown in Fig.~\ref{fig:s1}(c).  Impedance $Z_1$ represents the coupling between the CPW feedline and the nearest pad. Impedance $Z_4$ represents the cross capacitance between the feedline and the opposite pad, while $Z_5$ and $Z_3$ represent coupling of the pads to ground.  Impedance $Z_2$ represents the JJ array.

The circuit model in Fig.~\ref{fig:s2}(a) can be simplified by applying the standard Y-$\Delta$ transform, as shown in Fig.~\ref{fig:s2}(c). The resulting impedances $Z_1', Z_2', Z_5'$ are given by:
\begin{eqnarray}
  Z_1' &= \frac{Z_2Z_5 + Z_5Z_1 + Z_1Z_2}{Z_1} \\
  Z_2' &= \frac{Z_2Z_5 + Z_5Z_1 + Z_1Z_2}{Z_2} \\
  Z_5' &= \frac{Z_2Z_5 + Z_5Z_1 + Z_1Z_2}{Z_5}.
\end{eqnarray}
The total impedance of the structure is given by
\begin{equation}
  Z_{\rm total} = \frac{Z_1 Z_2 Z_3+Z_1 (Z_2+Z_3) Z_4+((Z_1+Z_2) Z_3+(Z_1+Z_2+Z_3) Z_4) Z_5}{Z_4 (Z_3+Z_5)+Z_1 (Z_2+Z_3+Z_5)+Z_2 (Z_3+Z_4+Z_5)}.
\end{equation}
We note that the frequency response of $Z_{\rm total}$ near resonance can be approximated using a simplified model as that shown in Fig.~1(d) of the main text.  Given the total impedance $Z_{\rm total}$ which shunts the CPW feedline, the scattering matrix can be directly computed using standard techniques \cite{Pozar}.

\section{Details on fabrication procedures}

The JJ arrays are fabricated by e-beam lithography at 100 kV using a Vistec 5000+ electron beam pattern generator. The use of a high energy electron beam minimizes forward scattering in the PMMA/MMA resist bilayer and allows implementation of a bridge-free double angle evaporation technique \cite{BFT}. We develop using a mixture 1:3 of IPA:water at 6 $^\circ$C. Prior to aluminum deposition, the substrate is cleaned of resist residues using a high pressure, low power oxygen plasma \cite{pop}. The aluminum films are deposited in a Plassys UMS300 UHV multichamber e-beam evaporation system with a base pressure of $10^{-9}$ Torr. The static oxidation in between the aluminum layer depositions is performed in a separate chamber, in a mixture 1:3 of $\mathrm{O}_{2}$:Ar at 100 Torr for 10 minutes. We obtain critical current densities of the order of tens of amperes per square centimeter. The junctions age by $\lesssim$10\% during the first week after fabrication.

\section{Modification of array mode frequencies by coupling pad capacitances}

\begin{figure}
  \includegraphics[width=0.7\textwidth]{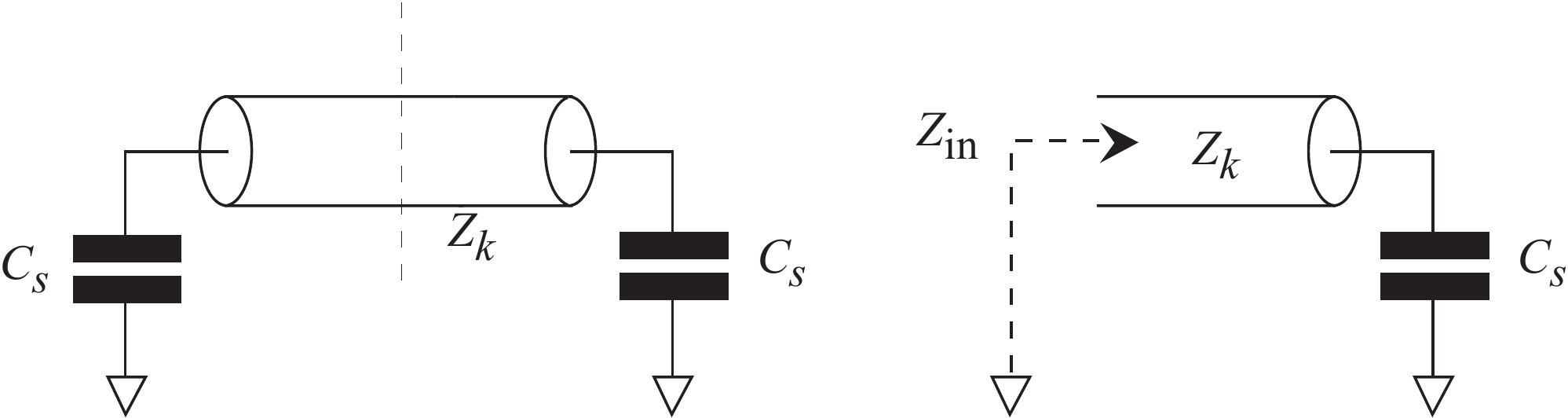}\\
  \caption{Transmission line model for $k$ modes of an $N$-junction array shunted by pad capacitances $C_S$.}\label{FigTL}
\end{figure}

The end capacitance pads lower the eigenmodes of the JJ array. To evaluate the modified mode frequencies of the $N$-junction array $\omega_{k}^{c};\; k\in [-N/2, N/2]$, we model each $k$ mode as a transmission line of characteristic impedance,
\begin{eqnarray}
    Z_{k} &=& \frac{1}{2}\sqrt{\frac{L_{k}}{C_{k}}} \nonumber\\
    &=& R_{q}\sqrt{\frac{2E_{C}/E_{J}}{\left(1 - \cos \frac{\pi k}{N}\right)\left(1 - \cos \frac{\pi k}{N} + \frac{C_{0}}{2 C_{J}}\right)}},
\end{eqnarray}
terminated by a shunt capacitance $C_{s}$ at both ends. Here $R_{q} = \hbar/ (2 e^{2}) = 1.02$ k$\Omega$ represents the reduced superconducting resistance quantum (Fig. \ref{FigTL}).
\par
\begin{figure}[htbp]
  \includegraphics[width=0.85\textwidth]{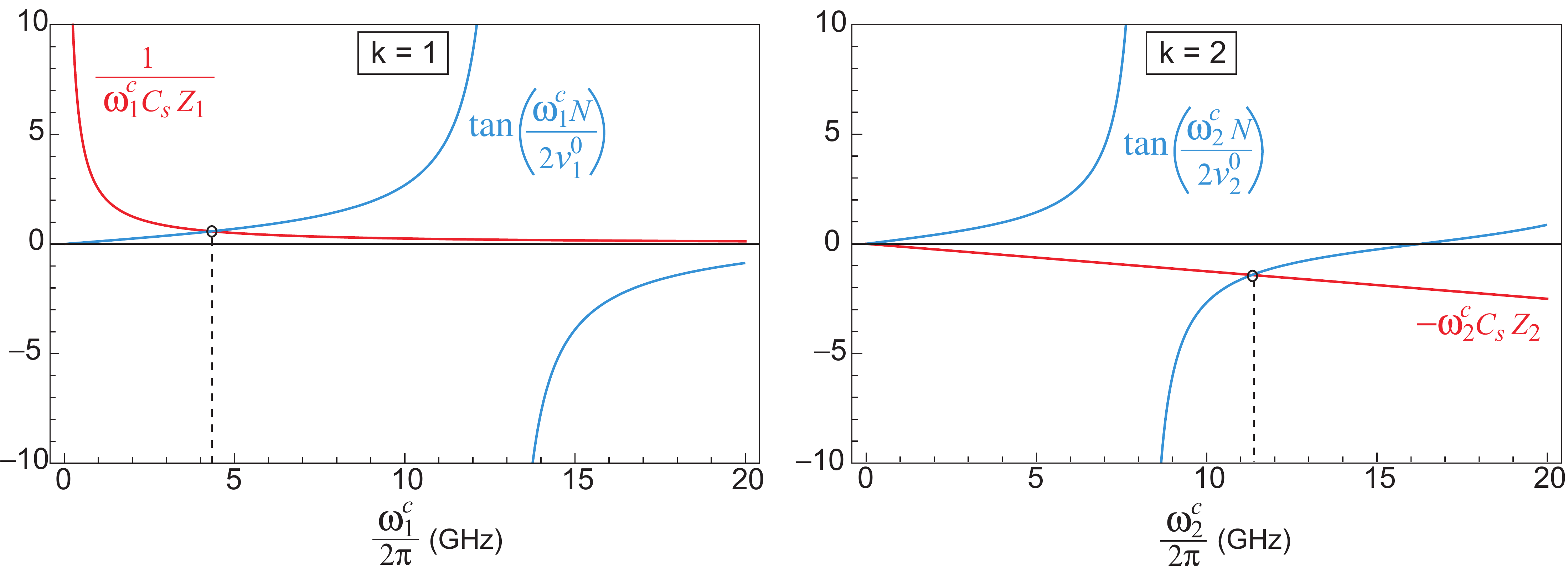}\\
  \caption{Determination of $f_{k=1}$ and $f_{k=2}$ modified by pad capacitances, using  Eqs.(\ref{treq}) and (\ref{treq2}) respectively. The fitting parameters were $C_{0}/C_{J}$ = 0.001, $C_{s}$ = 0.7 fF and $L_{J}$ = 1.9 nH.}\label{Fig:kplots}
\end{figure}
In order to calculate a mode frequency of the loaded transmission line, we exploit the symmetry of the problem by considering a half-section of the line.  The input impedance of each section can be written as \cite{Pozar}
\begin{eqnarray}
    Z_{\rm in} = Z_{k} \frac{Z_{L} + j Z_{k} \tan (\beta_{k} N/2) }{Z_{k} + j Z_{L} \tan (\beta_{k} N/2)},
\end{eqnarray}
where $Z_{L} = 1/ j \omega_{k}^{c} C_{s}$ and $\beta_{k}$ denotes the wave number associated with the $k^\mathrm{th}$ mode. At resonance frequency $\omega_{k}^{c}$, the impedances of left and right sections of the transmission line satisfy the identity $Z_{\rm in}^{\rm left}= [Z_{\rm in}^{\rm right}]^{*}$. Further, due to symmetry, the input impedances of the left and right sections of the transmission line are equal, $Z_{\rm in}^{\rm left}= Z_{\rm in}^{\rm right}$. These two conditions imply $Z_{\rm in} = 0$, which leads to a relationship,
\begin{eqnarray}
    \frac{1}{\omega_{k}^{c} C_{s} Z_{k}} =  \tan \left(\frac{\omega_{k}^{c}}{v_{k}^{0}}\frac{N}{2}\right).
    \label{treq}
\end{eqnarray}
Here $v_{k}^{0} = N\omega_{k}^{0}/(k \pi)$ denotes the phase velocity for each mode evaluated using the unloaded array calculation. Equation (\ref{treq}) is a transcendental equation in $\omega_{k}^{c}$, which can be solved numerically for each of the $k$ modes to obtain the corrected mode frequency. The above calculation yields the corrected mode frequencies $\omega_{k}^{c}$ only for the odd modes of the array. Similar treatment using the input admittance $Y_{\rm in}$ yields the corrected frequencies for the even modes, as solutions of the following transcendental equation 
\begin{eqnarray}
    -\omega_{k}^{c} C_{s} Z_{k} = \tan \left(\frac{\omega_{k}^{c}}{v_{k}^{0}}\frac{N}{2}\right).
    \label{treq2}
\end{eqnarray}
Fig.~\ref{Fig:kplots} shows example calculations for the first two mode frequencies using the method described above.

\bibliography{references}{}
\bibliographystyle{apsrev4-1}

\end{document}